\documentclass{ws-procs9x6}
\usepackage{amsmath}
\newcommand{\msum}{\sideset{}{'}\sum_{n=0}^{\infty}}

\newcommand{\re}{{\rm Re}}
\renewcommand{\imath}{\mathrm{i}}
\begin{document}

\title{Thermal effects in the magnetic Casimir-Polder interaction}

\author{H. HAAKH$^*$, F. INTRAVAIA%
\footnote[2]{Present 
address: Theoretical Division, MSB213, Los Alamos National Laboratory, Los Alamos, NM87545, USA.}%
, C. HENKEL}

\address{Institut f\"{u}r Physik und Astronomie, 
Universit\"{a}t Potsdam\\
 Karl-Liebknecht-Stra{\ss}e 24/25, 14476 Potsdam, Germany
\\
$^*$E-mail: haakh@uni-potsdam.de\\
www.quantum.physik.uni-potsdam.de}

\begin{abstract}
We investigate the magnetic dipole coupling between a metallic surface
and an atom in a thermal state, ground state and excited hyperfine state. 
This interaction results in a repulsive correction and -- unlike the electrical dipole contribution  -- depends sensitively on the Ohmic dissipation in the material.
\end{abstract}


\bodymatter

%===================================
\section{Introduction}
%===================================

Exact knowledge of the Casimir-Polder interaction between an atom and a
conducting surface is rapidly becoming important in modern
microtrap experiments (atom chips), whose stability is limited by an attractive
potential well at sub-micron distances.
%
%depends strongly on magnetic fluctuations. 
We investigate here the magnetic
dipole contribution to the atom-surface interaction 
and find it to be repulsive for atoms prepared in certain states. 
The magnetic Casimir-Polder potential differs greatly from
its electrical counterpart, on which previous research has concentrated.
Above all, the much smaller transition frequencies lead to a stronger dependence on temperature and the Ohmic dissipation in the material. This results in a strong suppression at distances above the thermal wavelength which is absent in
the case of the plasma and, more generally, in superconductors.
% of the magnetic dipole interaction in dissipative media.
%Furthermore, the interaction resembles the 
%in special configurations we could recover a non-vanishing
%entropy at T=0, in full analogy to the 
%Casimir interaction between
%two conducting plates.

%\section{Free energy of the Casimir-Polder interaction}

The fundamental quantity calculated in this work is the free energy of the magnetic dipole interaction between an atom and a planar surface.
For an atom prepared in a state $|a\rangle$ the free energy is given by 
the expression\cite{Wylie1985, Gorza2006}
\begin{multline}
\label{WS}
\mathcal{F}(L, T)=-k_{B}T \msum \beta^{a}_{ij}(\imath \xi_n)\mathcal{H}_{ji}(L,\imath \xi_n)
+\sum_b n(\omega_{ba}) \mu_i^{ab} \mu_j^{ba}
\re [\mathcal{H}_{ji}(L,\omega_{ba})]~,
\end{multline}
where $\xi_{n}=2\pi n k_{B}T/\hbar$ are the Matsubara frequencies and the
term $n=0$ comes with a weight $\frac12$ in the primed sum.
The second (resonant) term involves the mean thermal photon number
$n(\omega)$ and the magnetic transition dipole matrix $\boldsymbol{\mu}$.  
The 
% quantity $\boldsymbol{\beta}^a$ is the 
state-specific polarizability is
\cite{McLachlan1963,Wylie1985} 
\begin{equation}
%\begin{split}
\beta_{ij}^a({\omega})=\sum_b
\frac{\mu_i^{ab}\mu_j^{ba}}{\hbar}\frac{2\omega_{ba}}{\omega_{ba}^{2}-(\omega+\imath 0^{+})^{2}}~,
%\end{split}
\label{eq:WS_polarizability}
\end{equation}
while $\boldsymbol{\mathcal{H}}$ is the magnetic Green tensor
\begin{equation}
%\begin{split}
\boldsymbol{\mathcal{H}}(L,\omega) = \frac{\mu_0}{8 \pi}
\int\limits_0^\infty k dk \, \kappa
\left[\left(r_{\rm s}%(\omega, k)
+
 \frac{\omega^2}{c^2 \kappa^{2}}r_{\rm p}%(\omega, k)
 \right)[\boldsymbol{\hat x\hat x}+ \boldsymbol{\hat y\hat y}]
+ 
2\frac{k^{2}}{\kappa^{2}}r_{\rm s}%(\omega, k) 
\boldsymbol{\hat z\hat z}\right]e^{-2 \kappa L }~.
\label{eq:magnetic_GF}
%\end{split}
\end{equation}
Here, $\mu_{0}$ is the vacuum permeability,
$k$ is the in plane wave vector, 
$\boldsymbol {\hat x\hat x}$, $\boldsymbol{\hat y\hat y}$, $\boldsymbol{\hat z\hat z}$ are the cartesian dyadic products,
and $\kappa = [k^2 - (\omega + \imath 0^+)^2/c^2]^{1/2}$.
The Fresnel reflection amplitudes\cite{Jackson1975} 
$r^{\rm s,p}(k,\omega)$ are 
taken here for a local, isotropic and nonmagnetic
($\mu(\omega)=1$) medium.
%are given by
%
%\begin{equation}
%r^{\rm TE}(\omega, k)=\frac{\kappa-\kappa_{m}}{\kappa+\kappa_{m}},
%\quad
%r^{\rm TM}(\omega, k)=\frac{\epsilon(\omega)\kappa-\kappa_{m}}{\epsilon(\omega)\kappa+\kappa_{m}}~,
%\label{eq:fresnel}
%\end{equation}
%
%where $\kappa$, $\kappa_m$ are the propagation constants in vacuum
%and in the medium, respectively (roots with $\im{\kappa} \le 0$)
%
%\begin{equation}
%\label{eq:kappa}
%\kappa=\sqrt{k^2-\frac{\omega^2}{c^2}},\quad 
%\kappa_{m}=\sqrt{k^2-\epsilon(\omega)\frac{\omega^2}{c^2}},
%\end{equation}
%
%and $k=|\boldsymbol{k}|$ is the modulus of the in-plane wavevector.
%Note that the magnetic Green tensor can be obtained from the electric one 
%$\boldsymbol{\mathcal{G}}$ by 
%swapping the reflection coefficients \cite{Henkel1999}:
%\begin{equation}
%\label{eq:electric_GF}
%\boldsymbol{\mathcal{H}}\equiv c^{-2}\boldsymbol{\mathcal{G}}(r^{\rm TE}\leftrightarrow r^{\rm TM})
%\end{equation}
%
All information about the optical properties and the Ohmic dissipation in the surface is then encoded in the dielectric function $\varepsilon(\omega)$.
%We will use different commonly established descriptions, each of which includes Ohmic dissipation in a very characteristic way.
%As it turns out, the magnetic
%Casimir-Polder interaction is much more sensitive to dissipation than the
%electric one, because the magnetic transitions involve much smaller energies.
%
We will consider a \emph{Drude metal} \cite{Jackson1975}
%
%\begin{equation}
\(
\varepsilon_\mathrm{Dr}(\omega) =1 - \omega_{p}^{2} / [\omega(\omega+\imath\gamma)]
	\label{eq:def-Drude-epsilon}
\)
%\end{equation}
%
with plasma frequency $\omega_{p}$ and dissipation rate 
$\gamma > 0$, independent of temperature.
%This is the simplest model for a metal with finite conductivity. 
%If impurity scattering limits the charge mobility, $\gamma$ is constant and independent of temperature.
In the dissipationless \emph{plasma model},
we get $\varepsilon_\mathrm{pl}(\omega)$ 
% is obtained from $\varepsilon_{\rm Dr}(\omega)$ 
by setting $\gamma = 0$. It can be read as the limiting case of a  \emph{superconductor}\cite{Schrieffer1999} well below the transition temperature.
For a more thorough discussion of Casimir(-Polder) effects in superconductors, see 
Refs.\refcite{Bimonte2009b, Haakh2009b}.
%We adopt here (third model) a 
%description in terms of the two-fluid model, a weighted sum of a 
%dissipationless supercurrent response (plasma model) and a normal conductor
%response 
%â, which is particularly interesting for atom chips, as dissipation can be controlled by temperature.

%===================================
\section{Casimir-Polder potential in global equilibrium}
%===================================

%##########################################
\begin{figure}[bth]
\includegraphics[scale=.5]{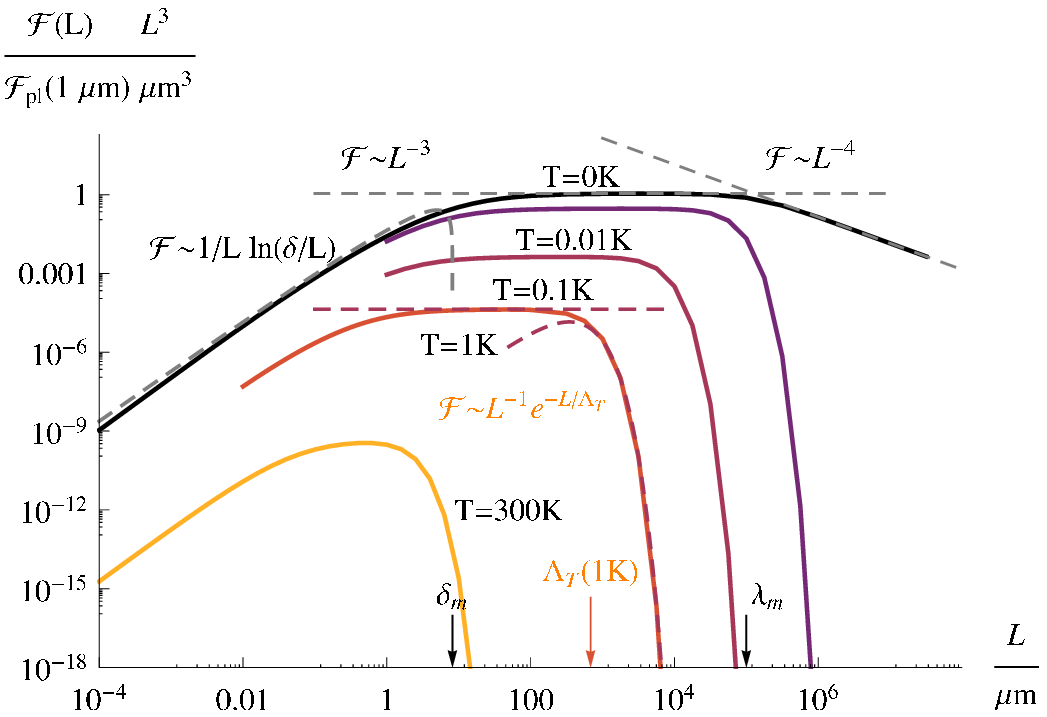}
\includegraphics[scale=.5]{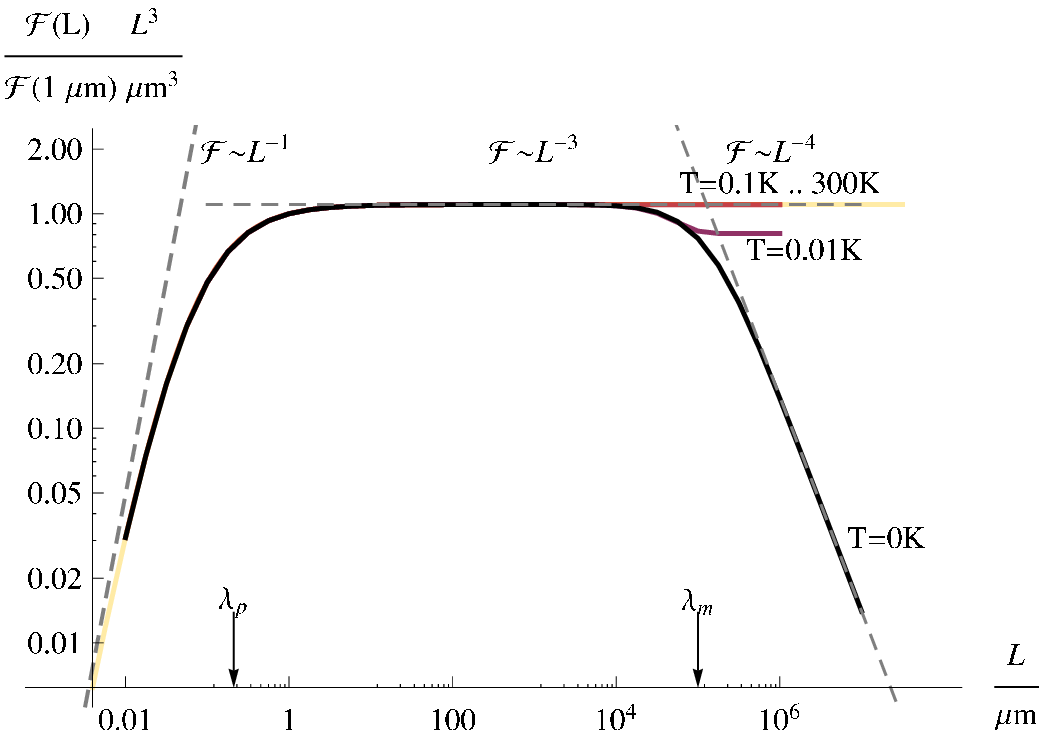}
\caption{Magnetic Casimir-Polder free energy for a thermalized
two-level atom near 
a Drude metal (left) and a plasma (right). 
Parameters are $\omega_p = 8.9 \times 10^{15} {\rm s}^{-1}, 
\gamma = 0.01 \omega_p, 
\Omega_m = 3\times10^9 {\rm s}^{-1} \approx 23 \, {\rm mK}$. 
The energy is scaled to $\mathcal{F}_{\rm pl}(1\,\mu{\rm m}, 0\,{\rm K}) 
= 9.79\times 10^{-37}\,\rm J$. 
Dashed asymptotes: see Ref. \refcite{Haakh2009b}.
}
	\label{fig:thermal}
\end{figure}
%##########################################

For an atom in a thermal state,
the polarizability Eq.\eqref{eq:WS_polarizability} must be averaged over thermal
state occupation numbers.
%
%\begin{equation}
%\beta^{T}_{ij}(\omega) = \sum_{a}
%\frac{e^{-E_{a} / k_B T}}{Z}
%\beta^{a}_{ij}(\omega)
%\label{eq:WS_polarizability_thermal}
%\end{equation}
%
%where $Z$ is the partition function. 
%In the limit $T\rightarrow 0$, we recover the polarizability for a ground state atom.
For a two-level system with transition frequency $\Omega_m$, this yields
\begin{equation}\label{eq:two-level_polarizability}
\boldsymbol{\beta}^{T}(\omega) 
= 
\tanh\left(\frac{\hbar\Omega_m}{2k_B T}\right) \boldsymbol{\beta}^{g}(\omega)
~,
\end{equation}
in terms of the ground state polarizability [Eq. (\ref{eq:WS_polarizability})
with $a=g$].
In thermal equilibrium, Eq.\eqref{WS} reduces to
\begin{equation}
\label{eq:Matsubara-series}
\mathcal{F}(L, T)=-k_{B}T
%{\rm Tr}
\msum
\beta^{T}_{i j}(\imath \xi_n)
%\cdot
\mathcal{H}_{j i}
(L, \imath \xi_n).
\end{equation}
Both $\boldsymbol{\beta}^{T}(\imath \xi)$ and $\boldsymbol{\mathcal{H}}(L, \imath \xi)$ are real expressions for $\xi>0$. We assume a static
magnetic dipole 
aligned perpendicular to the surface, not unrealistic in magnetic traps, and 
have $\beta^T_{xx} = \beta^T_{yy}$ and $\beta^T_{zz}=0$. 

The free energies for a Drude metal and a plasma are shown in Fig. \ref{fig:thermal}. A striking difference occurs at large distances: 
% The interaction of a magnetic dipole with 
the Drude metal is transparent to static
magnetic fields [$\boldsymbol{\cal H}( L, 0 ) = 0$],
% [$r^{\rm TE}(0)=0$]. 
and the zeroth term of the Matsubara sum vanishes. The free energy is then
dominated by the first term, which decays exponentially for 
$L > \Lambda_T =  \hbar c  / 4 \pi k_B T$. We call this the \emph{thermal decoupling} of the magnetic dipole.

A plasma (superconductor) shields static magnetic fields (Mei\ss{}ner-Ochsenfeld effect,  $\boldsymbol{\cal H}( L, 0 )  \ne 0$), 
leading to an enhanced interaction energy at nonzero temperature. The
linear dependence on $T$ of the zeroth Matsubara term
in Eq. \eqref{eq:Matsubara-series} cancels with the thermal 
polarizability (\ref{eq:two-level_polarizability}) for
$\hbar \omega_{ab} \ll k_B T$.
Only around this (quite low) temperature, there are any thermal effects.
% , and above one recovers the asymptotic for $T=0$ in the non-retarded regime
\cite{Haakh2009b}.
% strong differences in the magnetic atom-surface coupling.
The close coincidence between the thermal and the $T=0$ potentials
at small distances can be understood from the low-frequency
behaviour of the Green tensor, see Ref. \refcite{Haakh2009b} where also
the dashed asymptotes of Fig. \ref{fig:thermal} are discussed.
%Secondly, in the thermal regime, the interaction energy is enhanced with respect to $T=0$. Here, the free energy is given by the zeroth Matsubara summand  in Eq. \eqref{eq:Matsubara-series} where the linear temperature dependence cancels with the one of the polarizability (\ref{eq:two-level_polarizability}) due to the small transition frequencies $\hbar \omega_{ab} \ll k_B T$. 
 
 %##########################################
\begin{figure}[bth]
\includegraphics[scale=.5]{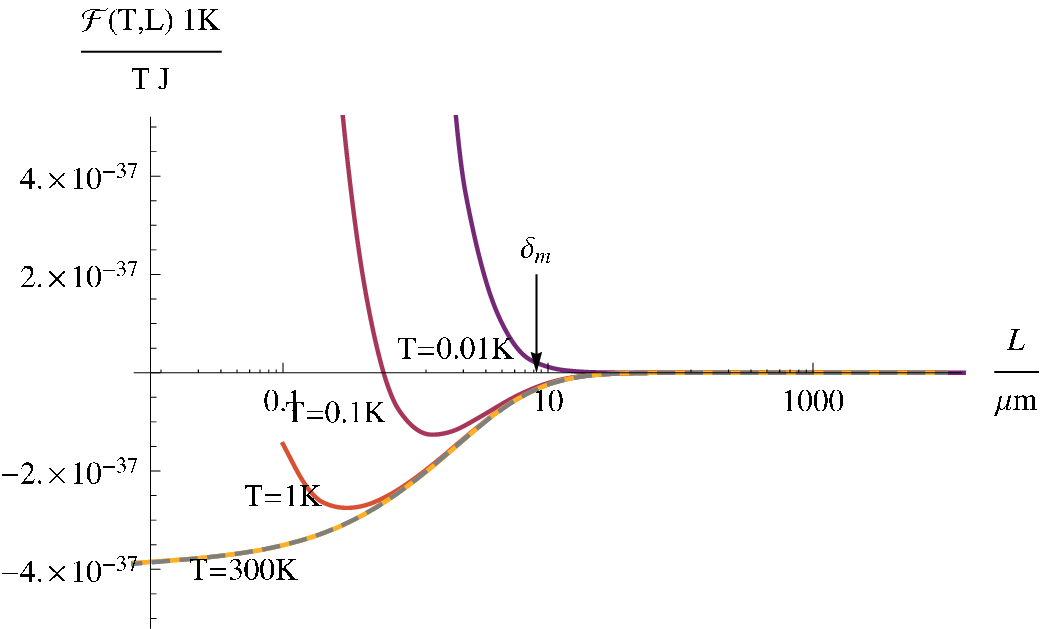}
\includegraphics[scale=.5]{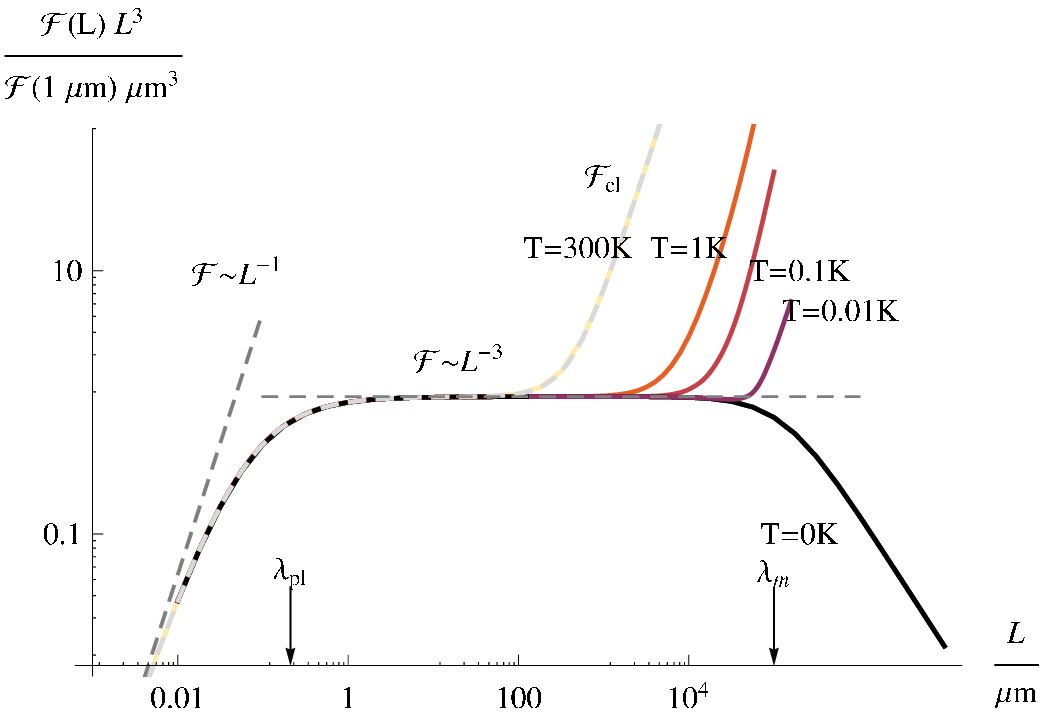}
\caption{Magnetic Casimir-Polder free energy in the atomic ground state
(left: Drude metal, right: plasma). Note the different scaling in the left plot.
Parameters as before.}
\label{fig:nonthermal}
\end{figure}
%##########################################

%===================================
\section{Casimir-Polder potential for nonthermal states}
%===================================
%
Many experimental settings involve atoms prepared in a special state rather than 
in thermal equilibrium with the environment. We consider a two-level atom 
prepared in the ground state as an example for an ultracold gas. The assumption
$k_B T \gg \hbar |\omega_{ba}|$ is quite realistic for most temperatures,
% the thermal occupation number $n(\omega_{ba})$ becomes linear.
and now the resonant part of Eq. \eqref{WS} contributes, too, and the free energy becomes 
\begin{eqnarray}
\label{WS2Level}
&&
\mathcal{F}(L, g, T)
%\approx-\frac{k_{B}T}{2}\beta^{0}(0)\mathcal{H}(L,0)\\
%+\frac{\hbar\Omega_{m}}{2}
%n(\Omega_{m}) \beta^{0}(0)\re[\mathcal{H}(L,\Omega_{m})]\\
\approx
- 
2 k_{B}T \sum_{n\ge 1}\beta^{g}(\imath\xi_{n})
\mathcal{H}_{xx}(L,\imath\xi_{n})
\\
&& {} 
\qquad 
+ k_{B}T\beta^{g}(0)\left\{
\re[\mathcal{H}_{xx}(L,\Omega_{m})]
-
\mathcal{H}_{xx}(L,0)
\right\}
\nonumber~.
\end{eqnarray}
At the temperatures considered, the first line (without the $n=0$ term) is nearly identical for the plasma and Drude model. Anyway, the second line is not. In the Drude model $\mathcal{H}(L,0)=0$, but the remaining resonant part is
significant in the non-retarded regime. It actually
changes the sign of the Casimir-Polder potential already at short
distances, as soon as $k_BT \gtrsim \hbar\Omega_m$,
see Fig. \ref{fig:nonthermal} (left). This leads to an \emph{attractive}
potential well of approximately $0.02\,\rm{pK}$ [environmental temperature
$T = 1\,\rm K$] at distances below $1\, \mu \rm m$, possibly accessible to quantum reflection experiments\cite{DeKieviet03}.
If we consider an atom prepared in an excited state (e.g. another hyperfine 
state),
% of $^{87}\rm{Rb}$), 
the interaction changes sign globally, because of the transition
energies $\hbar \omega_{ab}<0$ in the 
polarizability~\eqref{eq:WS_polarizability}.

In contrast, the magnetic coupling to a plasma is entirely repulsive at small distances (Fig.  \ref{fig:nonthermal}, right). Here, the second line of Eq.(\ref{WS2Level}) nearly vanishes because $\mathcal{H}_{xx}(L, \omega)$ is 
approximately independent of frequency,
at least in the non-retarded regime. 
Hence, the zeroth Matsubara term is removed from the Casimir-Polder
potential and the next order in the expansion of the occupation number 
$n( \Omega_m ) \approx k_B T / \hbar \Omega_m - \frac{ 1}{2 }$ 
gives the leading contribution to the resonant term.
In the non-retarded regime this restores
%
%\begin{equation}
%	\mathcal{F}^{\rm pl}_{\rm an}( L, g, T ) \approx
%	- |\mu_x|^2 \re \mathcal{H}_{xx}( L, \Omega_m )~,
	%\approx
	%\frac{ \mu_0 |\mu_x|^2 }{ 32\pi L^3 }
%	\label{eq:non-retarded-n-eq-plasma}
%\end{equation}
%
the $T = 0$ behaviour (cf. right panels of Figs. \ref{fig:thermal} and \ref{fig:nonthermal}). At larger distances (retarded regime) the 
%difference between the Green functions in the 
% second line of Eq.(\ref{WS2Level}) nonzero 
resonant term becomes dominant and oscillates with distance.  A scheme to enhance 
these oscillations for the electric Casimir-Polder interaction has been recently
proposed, using rovibrational states of polar molecules\cite{Simen}.

%===================================
\section{Discussion}
%===================================

At experimentally relevant temperatures, the magnetic Casimir-Polder interaction shows much richer effects than its electric counterpart. This is because the electric dipole coupling is dominated by electric fields which in all conductors are screened by surface charges at large scales. This masks any difference between
a normal metal and a superconductor (or the plasma model).
Differently, the boundary conditions for magnetic fields, relevant for the magnetic dipole, depend on surface currents and hence on Ohmic dissipation.
We have seen that including dissipation in the surface response leads to the thermal decoupling of the atom at distances beyond the thermal wavelength.
For atoms prepared in a nonthermal state, the balance between repulsive and attractive contributions can produce local extrema, whose sign is controlled
by the atomic (hyperfine) state. 
%We show that the sign of the magnetic coupling can be tuned by preparing atoms in a suitable state. 
Experimental tests of the Casimir-Polder
interaction may also answer remaining open questions on the temperature
dependence of the Casimir interaction, 
offering the advantage of a well-defined system that can be handled with high
precision.

%Due to the striking similarities with the Casimir interaction between conducting parallel plates, an experimental verification of the magnetic effects might thus shine new light also on Casimir interaction, that has been at the center of much interest lately. 

We acknowledge financial support by the European Science Foundation 
(ESF) within the activity `New Trends and Applications of the Casimir 
Effect',  %(www.casimir-network.com)
by the German-Israeli Foundation for Scientific Research and Development (GIF) %[HH]
 and by the Alexander von Humboldt foundation.% [FI].


\begin{thebibliography}{1}

\bibitem{Wylie1985}
J.~M. Wylie and J.~E. Sipe, {\em Phys. Rev. A} {\bf 32}, 2030 (1985).

\bibitem{Gorza2006}
M. Gorza and M. Ducloy, {\em Eur. Phys. J. D} {\bf 40}, 343 (2006).

\bibitem{McLachlan1963}
A.~McLachlan, {\em Proc. Roy. Soc. (London) A.} {\bf 271}, 387 (1963).

\bibitem{Jackson1975}
J.~D. Jackson, {\em Classical electrodynamics} (Wiley, New
  York, 1975).

\bibitem{Schrieffer1999}
J.~Schrieffer, {\em {Theory of Superconductivity}} (Perseus Books, 1999).

\bibitem{Bimonte2009b}
%G.~Bimonte, H.~Haakh, C.~Henkel and F.~Intravaia, 
%%Optical bcs conductivity with  arbitrary purity at imaginary frequencies 
%arXiv:0907.3775, (2009).
%[Bimonte08a]
G. Bimonte,
% ``Casimir effect in a superconducting cavity and the thermal controversy''
{\em Phys. Rev. A} {\bf 78}, 062101 (2008).

\bibitem{Haakh2009b}
H.~Haakh, F.~Intravaia, C.~Henkel, S.~Spagnolo, R.~Passante, B.~Power and
  F.~Sols, %{Temperature dependence of the magnetic Casimir-Polder interaction}
  submitted to \emph{Phys. Rev. A}, {preprint arXiv:0910.3133} (2009).

\bibitem{DeKieviet03}
% Experimental Observation of Quantum Reflection far from Threshold
V. Druzhinina and M. DeKieviet,
\emph{Phys. Rev. Lett.} {\bf 91}, 193202 (2003).

%\bibitem{Hinds1991}
%E.~A. Hinds and V.~Sandoghdar, {\em Phys. Rev. A} {\bf 43}, 398 (1991).

\bibitem{Simen}
S. Ellingsen, S. Buhmann and S. Scheel, 
%Enhancement of thermal Casimir-Polder potentials of ground-state polar molecules in a planar cavity 
{\em Phys. Rev. A} {\bf 80},  22901(2009).


\end{thebibliography}
\end{document}